\documentclass[a4paper,fleqn,usenatbib]{mnras}

\usepackage{newtxtext,newtxmath}
\usepackage{xcolor}
\usepackage[T1]{fontenc}
\usepackage{ae,aecompl}

\usepackage{graphicx}	% Including figure files
\usepackage{amsmath}	% Advanced maths commands
\usepackage{amssymb}	% Extra maths symbols

\defcitealias{Whitbourn2014}{WS14}
\title[Local Hole revisited]{Local Hole revisited: evidence for bulk motions and self-consistent outflow}

\author[T. Shanks et al.]{
T. Shanks$^1$,\thanks{E-mail: tom.shanks@durham.ac.uk (TS)}
L.M. Hogarth$^2$,
N. Metcalfe$^1$,
J. Whitbourn$^1$.\\
$^1$Department of Physics, Durham University, South Road, Durham, DH1 3LE, England\\
$^2$Department of Physics and Astronomy, University College London, Gower Street, London WC1E 6BT, England
}

\date{Accepted 2019... . Received 2019.......; in original form 2019 May 5}

\pubyear{2019}

\begin{document}
\label{firstpage}
\pagerange{\pageref{firstpage}--\pageref{lastpage}}
\maketitle

\begin{abstract}

We revisit our mapping of the `Local Hole', a large underdensity in the
local galaxy redshift distribution that extends out to redshift,
$z\approx0.05$ and a potential source of outflows that may perturb the
global expansion rate and thus help mitigate the present `$H_0$ tension'. First, we compare  local peculiar velocities
measured via the galaxy average redshift-magnitude Hubble diagram,
$\overline{z}(m)$, with a simple dynamical outflow model based on the
average underdensity in the Local Hole. We find that this outflow model
is in good agreement with our  peculiar velocity measurements from
$\overline{z}(m)$ and not significantly inconsistent with  SNIa peculiar
velocity measurements from at least the largest previous survey. This
outflow could cause an $\approx2-3$\% increase in the local value of
Hubble's constant. Second, considering anisotropic motions, we find that
the addition of the outflow model may improve the  $\overline{z}(m)$ fit of a bulk
flow where galaxies are otherwise at rest in the Local Group frame. We conclude
that the Local Hole plus neighbouring overdensities such as the Shapley
Supercluster may cause outflow and bulk motions out to
$\approx150$h$^{-1}$Mpc that are cosmologically significant and that
need to be taken into account in estimating Hubble's constant.

\end{abstract}


%%%%%%%%%%%%%%%%%%%%%%%%%%%%%%%%%%%%%%%%%%%%%%%%%%

%%%%%%%%%%%%%%%%% BODY OF PAPER %%%%%%%%%%%%%%%%%%

\section{Introduction}

There have been many studies of large scale structure of the Local
Universe. While the Local Group and the Local Supercluster on $\approx1$
and $\approx10$h$^{-1}$Mpc scales are now well established, there is
still  controversy over the possible existence of larger scale
structures (e.g. \cite{Watkins2009, Davis2011, Nusser2011, Macaulay2011,
Macaulay2012, Turnbull2012, Davis2014, Watkins2015, Nusser2014,
Nusser2016, Scrimgeour2016, Feix2017}). Evidence for bulk motions
originally suggested the presence of a `Great' or `Giant Attractor' either at
$\approx40$h$^{-1}$Mpc  or $\approx150$h$^{-1}$Mpc scales corresponding
respectively to the Hydra-Centaurus (see e.g. \cite{lyndenbell1988}) or
Shapley superclusters (see e.g. \cite{Mathewson1992, Lauer1992, Lauer1994, Tonry2000}) and
references therein. These studies generally use standard candles to
provide galaxy distances from which their `peculiar motions' can be
derived from their redshifts and sophisticated modelling procedures were
developed to compare the peculiar velocities and density fields to
derive cosmological parameters (see e.g.
\cite{Dekel1999,Carrick2015,Jasche2019}). In the later references the
earlier claims for bulk motions of significant amplitude become more
muted as the CMB evidence mounted for models with $\Omega_m\approx0.3$
rather than $\Omega_m=1$ (e.g. \cite{Spergel2003,Planck2018}). In
addition, a `Local Hole' has also been claimed particularly in the
Southern Galactic Cap (SGC) out to $\approx150$h$^{-1}$Mpc (see e.g.
\cite{Shanks1984,Frith2003,Busswell2004,Keenan2012,Whitbourn2014,
Whitbourn2016}). These latter studies are mainly based on observations
of galaxy clustering via galaxy counts in redshift surveys. Other
authors have also claimed the existence of smaller-scale local voids
e.g \cite{Tully2008,Tully2016, Rizzi2017, Pustilnik2019}.

%i.e. $60\la r \la 150$h$^{-1}$Mpc or $10.0<K<12.5$ when
%translated using  the homogeneous $\overline{z}(K)$ {\color{red}model}.

There are several other arguments supporting the idea that the Universe
may be more inhomogeneous than expected. In terms of the Local Hole
there is striking agreement between the galaxy number redshift distribution,  $n(z)$, of
\cite{Whitbourn2014} (\citetalias{Whitbourn2014}) and the galaxy cluster $n(z)$ from the
REFLEX II/CLASSIX  X-ray samples of \cite{Boehringer2015,
Boehringer2019} across the sky (see also Section \ref{sec:conclusions}). 

More theoretically, the SNIa and BAO Hubble diagrams produce evidence
for a cosmological constant which appears uncomfortably finely tuned and
this has led several authors to look for an escape route by
hypothesising a large local underdensity out to $z\approx0.4$, usually
modelled by a Lemaitre-Tolman-Bondi (LTB) non-Copernican cosmology (e.g.
\cite{Redlich2014}).  Others (e.g. \cite{Lukovic2019}) have used the
LTB approach with the more restricted aim of addressing the `$H_0$
tension'  between Cosmic Microwave Background
\citep{Planck2018} and local distance scale (e.g. \cite{Riess2018a,Riess2018b}
estimates of Hubble's Constant, $H_0$.
 
Many of the claims of bulk motions and local underdensities are unlikely
in the standard $\Lambda$CDM model. For example, \cite{Wu2017} suggest
that the likely amplitude of velocity fluctuation in the local
$\approx150$h$^{-1}$Mpc volume of \citetalias{Whitbourn2014}   is
$\la5$\% of what is needed to explain the current
difference between global and local $H_0$ estimates. Indeed,
\cite{Riess2018c} criticise the Local Hole velocity outflow model of
\cite{Shanks2019} on the same grounds, that its $\approx500$kms$^{-1}$
amplitude on $\approx100$h$^{-1}$ Mpc scales is unlikely under
$\Lambda$CDM at the $6\sigma$ level. \cite{Shanks2018} already argued
that on their reading of the \cite{Wu2017} and \cite{Odderskov2017}
papers, the significance under $\Lambda$CDM was lower, in the range
$1.9-3.9\sigma$. Of course, whether it is more plausible to appeal to
`new physics' outside $\Lambda$CDM to explain the $H_0$ tension rather
than the Local Hole can be debated.   We also note that other authors
emphasise that local underdensities compatible with $\Lambda$CDM at the
$\approx2\sigma$ level can at least partly explain the $H_0$ tension
(e.g \citealt{Wojtak2014}).

Here we return to consider the results of \citetalias{Whitbourn2014}  on
the underdensity and dynamics of the Local Hole. \cite{Shanks2019} used
a simple linear theory model to predict the outflow caused by this local
underdensity and, assuming that it was centred on our position, found
that it would make an $\approx2$\% reduction to the local distance scale
estimate of Hubble's Constant. \cite{Riess2018c,Kenworthy2019}
criticised the assumption that the underdensity was isotropic around our
position. Following \cite{Riess2018c}, these authors also
claimed that SNIa peculiar velocities from \cite{Scolnic2018} showed
that the effect of any local underdensity was lower than suggested by
\cite{Shanks2019}. In what follows we shall address both these issues, the
isotropy assumption in Section \ref{sec:past} and the SNIa results in
Section  \ref{sec:sn1a}. However, our main aim is to compare the outflow
model of \cite{Shanks2019} based on the Local Hole underdensity
estimates of \citetalias{Whitbourn2014} with the independent peculiar
velocity estimates of these authors and check for consistency (see
Section \ref{sec:observed}).

The structure of the paper is therefore as follows. In
Section \ref{sec:past} we first summarise the datasets and results used
by \citetalias{Whitbourn2014} and \cite{Whitbourn2016}; we also review
the evidence for the approximate isotropy of the `Local Hole' around our
position. Then, in Section \ref{sec:compvpec}, for the first time, we
directly compare the  peculiar velocities estimated by
\citetalias{Whitbourn2014}  via the  statistical Hubble
diagram, $\overline{z}(m)$, of \cite{Soneira1979} with the  outflow
velocity estimates from the dynamical model of \cite{Shanks2019}. We
further compare these  with peculiar velocities estimated from the
Pantheon SNIa survey of \cite{Scolnic2018} and the larger survey used by
\cite{Kenworthy2019}. We present our conclusions in
Section \ref{sec:conclusions}. Throughout we shall assume a cosmology
with $\Omega_\Lambda=0.7$, $\Omega_m=0.3$ and $H_0=100h$ km s$^{-1}$
Mpc$^{-1}$.

\section{Previous Datasets and Results}
\label{sec:past}

\citetalias{Whitbourn2014}  used 2MASS $K$ band photometry to $K<12.5$
to define the galaxy samples on which both their galaxy $n(z)$
distributions and peculiar velocity estimates were based. They worked in
three sky regions covering $\approx3000$deg$^2$ each for a total of
9161.7 deg$^2$ as given in their Table 2 and shown in their Fig. 1. They
also used the galaxy redshift surveys 6dFGRS for the two areas with
Declination $\delta<0^\circ$ and SDSS for the area with
$\delta>0^\circ$. The three areas are therefore termed 6dF-NGC, 6dF-SGC
and SDSS-NGC. Roughly speaking, SDSS-NGC is centred on the 
North Galactic Pole, 6dF-SGC is centred on the South Galactic Pole and
6dF-NGC is in the direction of the CMB dipole and the Shapley Supercluster.
\citetalias{Whitbourn2014}  first compared galaxy $n(z)$
distributions in these three areas with a homogeneous model based on an
assumed galaxy luminosity function (LF). 2MASS+GAMA
$K$-band galaxy counts were also used in consistency checks on the
normalisation used in the model LF to estimate the over- and
under-densities from the $n(z)$ at $K<12.5$. Underdensities were
detected in all three directions but most strongly in the 6dF-SGC
direction. These results were checked by \cite{Whitbourn2016} who used
maximum-likelihood methods to derive the density-redshift relations
independently of the LF and found these $n(z)$-based results to be
robust.

\begin{figure}
	\includegraphics[width=0.85\columnwidth]{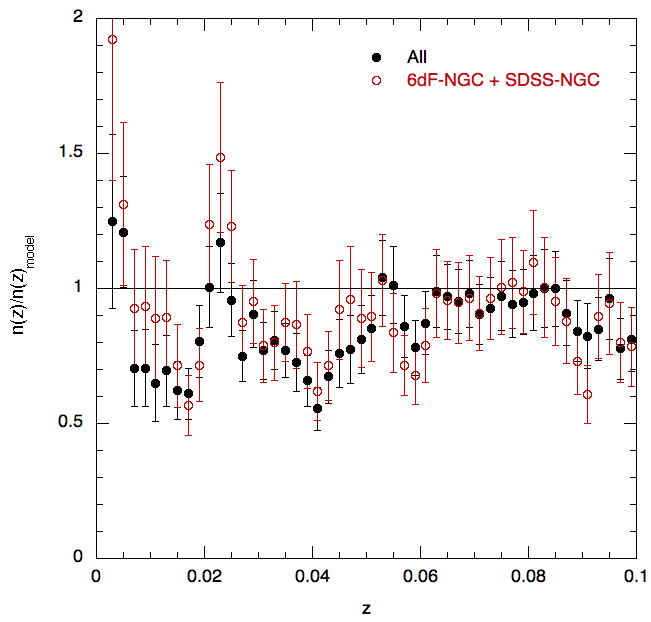}
    \caption{Density contrast-redshift relations for all three WS14  fields 
    and for 6dF-NGC$+$SDSS-NGC, each combined by area weighting. These both 
    show an underdensity out to $z\approx0.05$ implying that the underdensity
    is not just restricted to the 6dF-SGC area and that our assumption of an
    approximately isotropic `Local Hole' around our position is not unreasonable. 
    }
    \label{fig:nzall}
\end{figure}

The average galaxy redshift, $\overline{z}$, in $0.5$ mag
$K$  bins was then plotted versus $K$ magnitude in a Hubble diagram (see
Fig. 13 of \citetalias{Whitbourn2014}). 
\cite{Soneira1979} originally suggested using the statistic
$\overline{z}(m)$ to test the linearity of the Hubble law  in local
galaxy redshift surveys complete to some magnitude limit, in our case
$K<12.5$. Here, the $K$ band galaxy LF is implicitly assumed to be a
standard candle out to $z\la0.1$. It is relatively easy to predict
$\overline{z}(m)$ for the homogeneous case in the same way that galaxy
number-magnitude, $n(m)$, count models can be calculated.  Indeed, in
the case of a Euclidean model, the prediction is simply that
$\overline{z}(m)\propto10^{0.2m}$. The effects of cosmology,
K-correction and evolution are easily included in the model. The LF
normalisation $\phi^*(z)$ can also be used to eliminate the effects of
large-scale structure. The residual between the observed
$\overline{z}(K)$ Hubble diagram and the homogeneous model is then an
estimate of the peculiar velocity.

\citetalias{Whitbourn2014}  found that two directions showed evidence of
bulk motion ie galaxies at rest in the Local Group frame, while the
surveyed galaxies in the third 6dF-SGC direction were more consistent
with being at rest in the CMB frame. \citetalias{Whitbourn2014}
conjectured that the 6dF-SGC direction with its large underdensity might
be additionally affected by outflows which, if included, might improve
the fit of  bulk motion in the Local Group frame. We return to this
point in Section \ref{sec:observed}.

Finally, we shall also use the SNIa Pantheon survey of
\cite{Scolnic2018}. These include 1048 SNIa and  are the data used by
\cite{Riess2018a} to draw their Hubble diagram used to estimate $H_0$.
The same data plus additional unpublished Foundation+CSPDR3 SNIa surveys
was used by \cite{Kenworthy2019} to search for any velocity outflow
associated with the Local Hole. They concluded that the effect on $H_0$
was negligible. Here we use the 295 SNIa with $z<0.15$ in the Pantheon
sample of \cite{Scolnic2018} to compare  with the $\overline{z}(m)$
peculiar velocities of \citetalias{Whitbourn2014}. We note that the
statistical precision will be reduced by the loss of the
Foundation+CSPDR3 surveys' 102 SNIa with $0.023<z<0.15$. Also, both SNIa
surveys have very non-isotropic sky coverage (see Fig. 3 of
\cite{Kenworthy2019}). This leaves  only 6 SNIa in 6dF-NGC, 29 in
6dF-SGC and 20 in SDSS-NGC in the required $0.02<z<0.05$ redshift range
in the Pantheon survey. But we shall also compare these
results to those from the full  Pantheon+Foundation+CSPDR3 survey as
reported by \cite{Kenworthy2019}.

\begin{figure}
	\includegraphics[width=0.85\columnwidth]{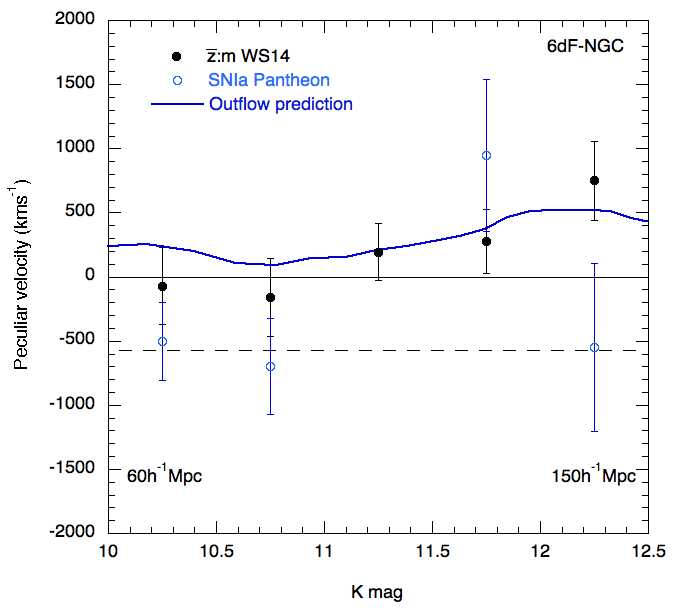}
    \includegraphics[width=0.85\columnwidth]{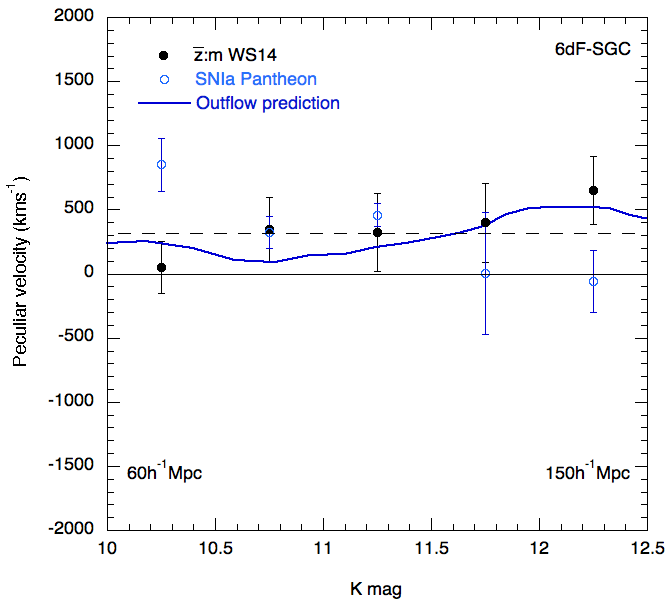}
	\includegraphics[width=0.85\columnwidth]{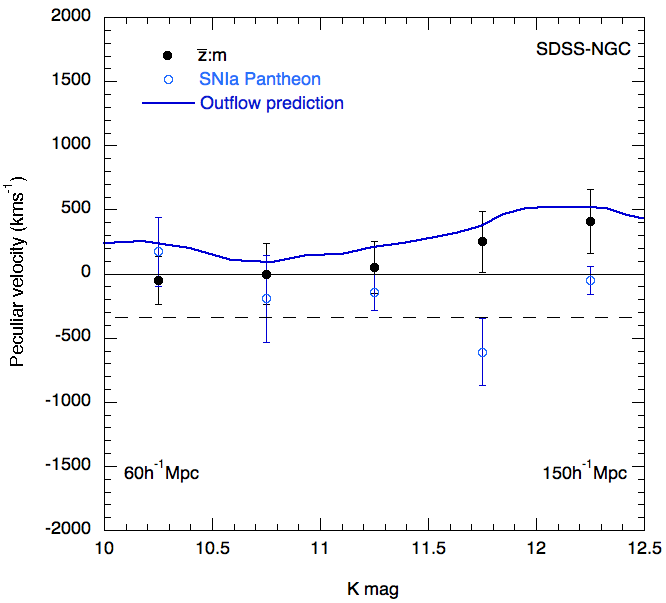}
   \caption{The WS14  peculiar velocities (filled circles)
   estimated from the residuals between the observed $\overline{z}(m)$
   Hubble diagram and a homogeneous model  in their three individual
   fields. The range $10.0<K<12.5$ translates to $60\la d \la
   150$h$^{-1}$Mpc via this model. WS14 found that bulk motion in the
   Local Group frame (solid horizontal line) was preferred by these data except in
   6dF-SGC where galaxies appeared more at rest in the CMB frame (horizontal dashed
   line). Adding the Local Hole outflow model of Shanks et al (2019) in
   the Local Group frame (blue line) improves the fit in 6dF-SGC while
   maintaining it in 6dF-NGC and SDSS-NGC. The peculiar velocities from
   Pantheon SNIa (open circles) show less good agreement with the model at larger
   distances.}
   \label{fig:vpec}
\end{figure}

%\section{Observed and predicted $v_{pec}$}
\section{Observed vs. predicted outflows}
\label{sec:compvpec}

\subsection{Predicted outflow model}
\label{sec:model}

\cite{Shanks2019} based their predicted `Local Hole' outflow model
on a simple linear theory gravitational growth model based 
on an assumed isotropic local galaxy underdensity as follows:

\begin{equation}
\frac{\Delta v}{v_H}=-\frac{1}{3}\frac{\delta\rho_g(<r)}{\bar{\rho}_g} \frac{\Omega_m^{0.6}}{b}
\label{eq:linear}
\end{equation}

\noindent where $\Delta v$ is the peculiar velocity at Hubble velocity,
$v_H$, corresponding to comoving radius, $r$, and $b$ is the galaxy bias.
$\delta\rho_g(r)/\bar{\rho}_g$ is the density contrast given by

\begin{equation}
\frac{\delta\rho_g(<r)}{\bar{\rho}_g}= \frac{1}{V(r)}\sum_{i} \bigg(\frac{dn}{n}\bigg)_i 4\pi r_i^2\delta r.
\label{eq:density}
\end{equation}

\noindent where $\big(\frac{dn}{n}\big)_i$ are taken from averaging the
data shown in Fig. 3 (a, b, c) of \citetalias{Whitbourn2014}. $r_i$ are the corresponding
comoving distances, $\delta r$ is the comoving bin size and $V(r)$ is
the spherical volume to radius, $r$. Clearly it is the $4\pi$ factor in
eq. \ref{eq:density} that represents our assumption that the 
$\big(\frac{dn}{n}\big)_i$ apply isotropically  over the whole sky.

Since \citetalias{Whitbourn2014}  only showed the individual $n(z)$'s for their three  areas, 
for completeness we first show in Fig. \ref{fig:nzall} the overall
average density contrast $\big(\frac{dn}{n}\big)_i$ found by combining
the three areas  of \citetalias{Whitbourn2014}  that leads to the $\frac{\Delta v}{v_H}(z)$
result shown in Fig. 1 of \cite{Shanks2019}. Here, we find an overall
median underdensity of $\approx-23$\% out to $\approx150$h$^{-1}$Mpc.
Indeed, also from Fig. \ref{fig:nzall}, the similarly combined SDSS-NGC
$+$ 6dF-NGC areas still show a $\approx-11$\% median underdensity.
Together, these results therefore support a roughly isotropic underdensity around
our position as assumed in the above model, now shown as the solid blue lines 
in Figs. \ref{fig:vpec}, \ref{fig:vpecall}.

\subsection{Observed $\overline{z}(m)$ outflows and bulk motions}
\label{sec:observed}

\begin{figure}
	\includegraphics[width=\columnwidth]{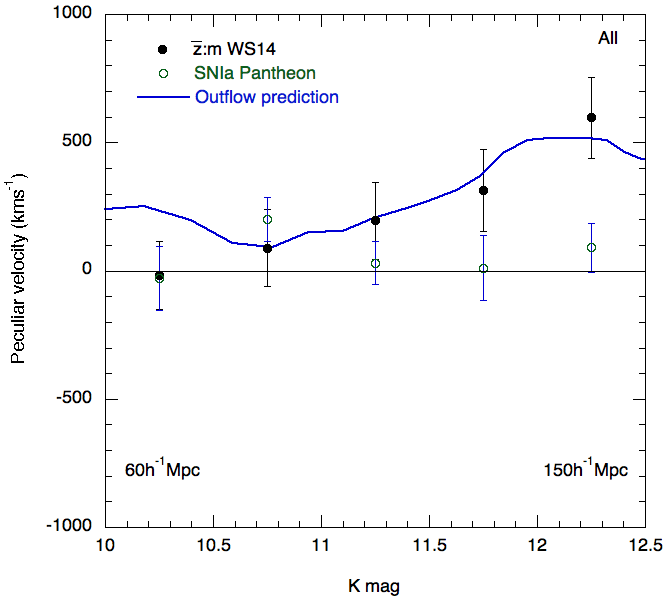}\vspace{-0.5cm}
    \caption{Now averaging over the three directions from Figs.
    \ref{fig:vpec}, the overall peculiar velocities from
    $\overline{z}(m)$ (filled circles) show excellent agreement with the
    Local Hole outflow prediction of Shanks et al (2019) (blue line), here added to
    the model where all galaxies are assumed to be at rest
    in the Local Group frame (horizontal solid line). The peculiar
    velocities from Pantheon SNIa (open circles) show less good
    agreement at larger distances/magnitudes. 
   }
   \label{fig:vpecall}
\end{figure}

In Figs. \ref{fig:vpec}  we summarise the \citetalias{Whitbourn2014} 
peculiar velocity results in each of their three areas, as estimated via
$\overline{z}(m)$. The closed circles show the residuals from the
homogeneous $\overline{z}(m)$ model which represent the
\citetalias{Whitbourn2014}  average peculiar velocity estimates in their
three directions. The homogeneous $\overline{z}(m)$ model has also been
used to relate the $K$ magnitude of a $\overline{z}(m)$ bin to its
luminosity distance, $d$. So the $K=10.25\pm0.25$ bin corresponds to
$d\approx60$h$^{-1}$ Mpc  and the $K=12.25\pm0.25$ bin corresponds to
$d\approx150$h$^{-1}$ Mpc, as indicated in Figs. \ref{fig:vpec},
\ref{fig:vpecall}. \citetalias{Whitbourn2014}  corrected their galaxy
redshifts into the Local Group frame (see \citetalias{Whitbourn2014} 
eq. 10) which implies the Local Group is moving with 633 kms$^{−1}$ with
respect to the Cosmic Microwave Background. 

Fig. \ref{fig:vpec} thus shows the $\overline{z}(m)$ peculiar velocities
compared to the $v_{pec}=0$ solid horizontal line that corresponds to galaxies
lying at rest in the Local Group frame whereas the horizontal dashed
line corresponds to the galaxies lying at rest in the CMB
frame. Thus the former would indicate the Local Group and galaxies in
that direction were participating in coherent bulk motion relative to
the CMB. \citetalias{Whitbourn2014}  concluded that the results in the
6dF-NGC and SDSS-NGC directions were consistent with such a bulk motion while
those in the 6dF-SGC direction were more consistent with the galaxies
being at rest in the CMB frame. They noted that the 6dF-SGC result might
still be consistent with bulk motion if there was an additional outflow
component due to the enhanced underdensity in that direction.

In Figs. \ref{fig:vpec} we now  compare the results for the above Local
Hole outflow model of \cite{Shanks2019} to the $\overline{z}(m)$
peculiar velocity results of \citetalias{Whitbourn2014}. 
The model (solid blue line) as plotted in these Figures simply
represents an adding of the outflow model to the $v_{pec}=0$ (solid
horizontal line) result expected if the galaxies are  at rest in the
Local Group frame. In the three areas this  combined bulk
flow plus outflow model looks consistent with these data. Thus the
addition of the outflow model seems to have improved the bulk motion fit
in the 6dF-SGC direction while not damaging the bulk motion fit too much
in the other two directions.

%Thus the assumption that the Local Hole is approximately symmetric around 
%our position at least appears consistent with the $\overline{z}(m)$ data.
%and overall in Fig. \ref{fig:vpecall} the model and peculiar velocity
%results are in  good agreement

% Omega_m and h varies using zLG from pantheon_zcmb_zLG_ra_dec.txt
%19.254     4           6   73.70+-0.2  0.28+-0.015 1.0558 1046 0*dv p
%19.254     2           6   72.70+-0.2  0.32+-0.01  1.0593 1046 1*dv p

\begin{table*}
\begin{center}
\begin{tabular}{cccccc}
\hline
         &  $\overline{z}(m)$ $\chi^2$ (d.f.) & $\overline{z}(m)$ $p$ &   SNIa $\chi^2$ (d.f.)   & SNIa $p$ & No. SNIa   \\
\hline
6dF-NGC   & 2.3778 (5)      & 0.79     & 0.6574 (4)      &  0.96                              &	 6\\
6dF-SGC   & 1.7763 (5)      & 0.88     & 18.114 (5)      &  2.8$\times10^{-3}$  ($3.0\sigma$) &  29\\ 
SDSS-NGC  & 2.5244 (5)      & 0.77     & 26.491 (4)      &  2.5$\times10^{-5}$  ($4.2\sigma$) &  20\\
All       & 2.4278 (5)      & 0.79     & 18.829 (5)      &  2.1$\times10^{-3}$  ($3.1\sigma$) & 295\\
\hline
\end{tabular}
\end{center}
\vspace{-0.4cm}
\caption{$\chi^2$ comparisons of the Local Hole outflow model of 
\protect\cite{Shanks2019} with $\overline{z}(m)$ and SNIa peculiar
velocity estimates. d.f. gives $\chi^2$ degrees-of-freedom. $p$ is the probability
of a higher $\chi^2$ value and $\sigma$  is the equivalent
2-tailed Gaussian significance level.}
\label{table:chi2}
\end{table*}

Fig. \ref{fig:vpecall} then shows the area weighted average
of the observed peculiar velocities over all  three directions. There is
clearly now no sensitivity to bulk motion but these data can still be
used to look for an outflow due to a global underdensity in this
volume. We again see excellent agreement between the
outflow model and the $\overline{z}(m)$ peculiar velocity estimates.

We have used $\chi^2$ to compare the $\overline{z}(m)$ peculiar
velocities with the predicted outflow model in each of the directions
and then in the three directions combined (see Table \ref{table:chi2}).
We have assumed the errors on the outflow model from \cite{Shanks2019}.
There are caveats on our application of these $\chi^2$ tests which we
discuss in Section \ref{sec:sn1a} below. Nevertheless, we see that the
$\chi^2$ values for the $\overline{z}(m)$ velocities appear generally
consistent with the model.

The observed \citetalias{Whitbourn2014}  peculiar velocities therefore
seem to  agree with the  outflow model of \cite{Shanks2019} based on the
\citetalias{Whitbourn2014}  Local Hole underdensity. Again it must be
cautioned that much depends on the final $K=12.5$,
$r\approx150$h$^{-1}$Mpc $v_{pec}$ point that \citetalias{Whitbourn2014}
regarded as uncertain, partly due to its amplitude. We also refer to
\citetalias{Whitbourn2014} 's caveat about the possible vulnerability of
 $\overline{z}(m)$ to evolution in the LF, although this is minimised by
the low redshift range involved and our use of the $K$ band. {\it
Nevertheless, in this work we recognise for the first time the
self-consistency of the \citetalias{Whitbourn2014}  Local Hole
underdensity and peculiar velocity measurements, related through our dynamical
outflow model.}

\subsection{SNIa peculiar velocities compared to outflow model}
\label{sec:sn1a}

We next compare the agreement between the peculiar velocities implied by
SNIa and our outflow model. In Figs. \ref{fig:vpec}, we see that the
agreement between the SNIa Pantheon results and the outflow model 
data   is poor in the fields with most SNIa ie the 6dF-SGC and SDSS-NGC
fields. In 6dF-NGC there is no inconsistency between these two but with
only 6 SNIa the errors are large. Also note that there are zero SNIa in
the $K=11.25$ bin. Clearly, the agreement with the model appears poorer for the
SNIa than the $\overline{z}(m)$ peculiar velocity estimates.

The all-sky results for 130 Pantheon SNIa in the usual magnitude range
$10<K<12.5$ or $60\la d \la 150$h$^{-1}$Mpc are shown in Fig.
\ref{fig:vpecall} where they are seen to disagree with the Local Hole
outflow prediction and the $\overline{z}(m)$ peculiar velocity estimates
of \citetalias{Whitbourn2014}. This result at first looks similar to that of
\cite{Kenworthy2019} who found no evidence of outflow in
the Pantheon $+$ Foundation $+$ CSPDR3 datasets.

We again use $\chi^2$ to compare the SNIa peculiar velocities with the
predicted outflow model in each of the three \citetalias{Whitbourn2014}
directions and then in the three directions combined (see Table
\ref{table:chi2}). We again assume the errors on the outflow model from
\cite{Shanks2019}. We see that the $\chi^2$ values for the the SNIa
peculiar velocities reject the model for the full sample and for the
SDSS-NGC and 6dF-SGC sub-samples and are generally poorer fits to the
model than the $\overline{z}(m)$ velocity estimates.

However, as previously noted, there are several issues which mean all
the $\chi^2$ results in Table \ref{table:chi2} should only be treated as
illustrative. First, we have ignored SNIa systematics which will
contribute to the errors. Second, we have ignored covariances  between
the $\overline{z}(m)$ bins,  the SNIa bins and the outflow model
predictions. The results are also sensitive to how we have handled the
errors on the model velocities in the $\chi^2$. All we claim is that
since we have applied our assumptions consistently to both the
$\overline{z}(m)$ and SNIa comparisons, so the {\it relative}
goodness-of-fits given in Table \ref{table:chi2} may be qualitatively
believable and useful. With these caveats, we conclude
that  the outflow model appears in better agreement with the
\citetalias{Whitbourn2014} $\overline{z}(m)$ peculiar velocities  than
with the Pantheon SNIa data.

The lack of detection of outflow is confirmed by our analysis of the
Pantheon SNIa survey over its full redshift range.\footnote{The 
caveat made above about ignoring the
systematic errors in the SNIa peculiar velocity estimates also applies
here.} Here with 1048 SNIa we found a reduced $\chi^2=1.0558$ on 1046
degrees-of-freedom for the best fit $\Omega_m=0.28\pm0.015$,
$H_0=73.7\pm0.2$kms$^{-1}$Mpc$^{-1}$ model with no outflow assumed.
Assuming the isotropic outflow model predicted by our Local Hole
results, we found a reduced $\chi^2=1.0593$ on 1046 degrees-of-freedom
for the best fit $\Omega_m=0.32\pm0.015$,
$H_0=72.7\pm0.2$kms$^{-1}$Mpc$^{-1}$ model. So the outflow model with
slightly higher $\Omega_m$ and slightly lower $H_0$ fits the Hubble
diagram less well but only by $\Delta \chi^2=3.66$ which, with only two
parameters fitted, represents only a $\approx1.4\sigma$ rejection of the
outflow model.\footnote{Note that here we have fitted in
the Local Group frame whereas \cite{Shanks2019} fitted in the CMB frame
after correcting for peculiar motions estimated from 2M++
\citep{Carrick2015} and found that the outflow model gave a marginally
better fit than no outflow.} This result broadly supports the reply of 
\cite{Shanks2019} to \cite{Riess2018c}, that a small increase in 
$\Omega_m$ allows our Local Hole outflow model to be an acceptable fit to the
SNIa Hubble diagram.

%$ prob (\Delta \chi^2>3.66$)=16\%

% for all 3 directions zbarm, chisq=2.4278 on 5df
% for all 3 directions SNIa, chisq=18.829 on 5df 2.9sigma
% 6dF-NGC zbarm  chisq=2.3778 on 5df
% 6dF-NGC SNIa  chisq=0.65736 on 4df
% 6dF-SGC zbarm  chisq=1.7763 on 5df
% 6dF-SGC SNIa  chisq=18.114 on 5df
% SDSS-NGC zbarm  chisq=2.5244 on 5df 77%
% SDSS-NGC SNIa  chisq=26.491 on 5df

Finally, \cite{Kenworthy2019} claim that in their bigger sample of 397
$0.023<z<0.15$ SNIa, our outflow prediction can be rejected in the
all-sky  case and in the particular  \citetalias{Whitbourn2014}  directions. \cite{Shanks2019}
assumed $z=0.1$ as being typical of a SNIa sample used to estimate $H_0$
at $z<0.15$, giving  $v_{pec}=540$kms$^{-1}$ from their Fig. 2 and
$v_{pec}/cz=1.8$\%. Now, this outflow prediction of $\Delta
H_0=H{_0}^{local}/H{_0}^{global}=1.018$ translates to a change in their
Hubble diagram intercepts of $\Delta a_B =a_B-a_B^{FLRW}=log_{10}(\Delta
H_0)=0.0077$. Based on the all-sky fits in their Table 1 (and Fig. 3a),
$\Delta a_B$=0.0077 is only rejected at $1.9-2.2\sigma$ in the
$0.023<z<0.15$ and $0.01<z<0.5$ ranges. Similarly, in their Table 1 (and
Fig. 5b), $\Delta a{_B}^{z<0.05}$=0.0077 is only rejected at $0.8\sigma$
in the $0.01<z<0.5$ range in the \citetalias{Whitbourn2014}  fields. \cite{Kenworthy2019} here
quote a $2.6\sigma$ rejection taking $v_{pec}=520$kms$^{-1}$ at $z=0.05$
from Fig. 2 of \cite{Shanks2019}. But volume weighting the outflow model
gives 365kms$^{-1}$ at $z\approx0.04$ leading to a $2.0\sigma$
rejection. So the rejection of the outflow model of \cite{Shanks2019} is
only at the $\approx1-2\sigma$ level and thus perhaps less strong in the larger
local SNIa samples of \cite{Kenworthy2019} than in the Pantheon
sub-sample.

\section{Conclusions}
\label{sec:conclusions}	

\citetalias{Whitbourn2014}  presented strong evidence from both galaxy
counts and galaxy number redshift distributions for a local
inhomogeneous underdensity out to $\approx150$h$^{-1}$Mpc. This `Local
Hole' underdensity was somewhat more pronounced in the 6dF-SGC (SGP) direction
but we have first shown here that the underdensity persists
after averaging over all three directions (see Fig. \ref{fig:nzall}).
In the Southern sky, the underdensity shown by the galaxy redshift
distribution is further strongly supported by the redshift distribution of
REFLEX II X-ray galaxy clusters (see Fig. 8 of \citealt{Boehringer2015}).
\cite{Boehringer2019} have recently also similarly found
that their new Northern cluster samples are in good agreement with the
\citetalias{Whitbourn2014}  $n(z)$ results. Combined, their CLASSIX
cluster survey covers 66\%  of the sky. Thus despite the criticism by
\cite{Kenworthy2019} of \citetalias{Whitbourn2014}  areas only covering
$\approx22$\% of the sky, this agreement with, and extra coverage of,
the X-ray cluster survey supports the possibility that the Local Hole
may feature over most of the local volume out to
$\approx150$h$^{-1}$Mpc. This further motivates the  assumption of
approximate isotropy made in the dynamical outflow model of
\cite{Shanks2019} (see also \cite{Hoscheit2018}). 

We note that \cite{Jasche2019} failed to detect a Local Hole
underdensity in 2M++ data. Nevertheless their Figs. 10(a, b, c) show
some similarity to our Fig. \ref{fig:nzall} and  Figs. 3(a, b) of
\citetalias{Whitbourn2014} but perhaps with different normalisations.
Although sophisticated in their treatment of peculiar velocities, in
deriving density-redshift relations \cite{Jasche2019} assume an LF
independent of galaxy colour and morphology unlike
\citetalias{Whitbourn2014}  and with no attempt to solve for the LF
independently like \cite{Whitbourn2016}. It would also be interesting to
check that their assumed LF and normalisation used to estimate their
density-redshift relations are consistent with fainter $K$-band galaxy
counts (c.f. Figs. 5, 6 of \citetalias{Whitbourn2014}).

%At $z\approx0.02$, $0.04$ in their Fig 10a, 
%$n(z)$ features have also clearly  been moved by their peculiar velocity
%corrections.

Overall, in Fig. \ref{fig:vpecall} the outflow model of
\cite{Shanks2019} also seems to fit the $\overline{z}(m)$ data well.
However, there is a discrepancy with the SNIa data from the Pantheon
sample which would prefer zero outflow as the best fit.  But we have
argued that the larger Pantheon+Foundation+CSPDR3 SNIa sample of
\cite{Kenworthy2019} only excludes the  void at the $1-2\sigma$ level.
Of course, there are still possible issues with the $\overline{z}(m)$ results:
they need substantial correction for the same local inhomogeneities that
are the subject of the $n(m)$ and $n(z)$ studies; they assume that there
are no evolutionary or environmental effects on the $K$-band LF used as
a standard candle. The highest redshift $\overline{z}(m)$ is also most
sensitive to systematic effects. {\it But the agreement between the
observed $\overline{z}(m)$ and the $v_{pec}$ outflow model is
impressive, adding to the strong evidence for the Local Hole from
the basic count and redshift survey data of \citetalias{Whitbourn2014}.}

In terms of  anisotropic flows, \citetalias{Whitbourn2014}  originally found that in two
directions the $\overline{z}(m)$ peculiar velocities implied the
galaxies were exhibiting bulk motions out to $\approx150$h$^{-1}$Mpc in
the sense that the galaxies appeared to be moving coherently with the
Local Group. In the other 6dF-SGC direction, that showed the biggest
underdensity, the result was more consistent with galaxies being at rest
in the CMB frame with no bulk motion. The suggestion by \citetalias{Whitbourn2014}  that the
addition of an outflow component might improve the agreement with the
bulk flows found in the other two directions now seems to be supported
by the current outflow model. While  improving the fit of the
$\overline{z}(m)$ peculiar velocities to the bulk motion solution in the
6dF-SGC direction the model maintains the bulk flow solution in the
other two  directions.

We conclude that an outflow component due to the Local Hole coupled with
a bulk motion within  an $\approx150$h$^{-1}$Mpc radius in the direction
of motion of the Local Group  towards the Shapley supercluster give a
self-consistent description of the \citetalias{Whitbourn2014}  density
and velocity fields implied by $n(z)$ and $\overline{z}(m)$ statistics. The size of
the resulting reduction in $H_0$ is at the $\approx2-3$\% level needed to
reconcile the reduced `tension' between the value of $H_0=67.4\pm1.7$km
s$^{-1}$ Mpc$^{-1}$ of \cite{Planck2018} and at least the $H_0=69.8\pm1.9$ km
s$^{-1}$ Mpc$^{-1}$ estimated from the TRGB distance scale of
\cite{Freedman2019}. The reasons for the discrepancy with the Pantheon
SNIa results are unclear but we have argued there is less disagreement
with the bigger Pantheon+Foundation+CSPDR3 SNIa survey as used by
\cite{Kenworthy2019}. It will be interesting to see how the SNIa results
improve at least in the Southern Hemisphere when more isotropic and
better sampled SNIa searches start with LSST in the next few years.

\section*{Acknowledgements}
We thank D. Scolnic (Duke University, USA) for supplying full information for the Pantheon SNIa sample.
We thank H. B{\"o}hringer (MPE, Germany) for informing us of the Northern Hemisphere CLASSIX X-ray cluster results prior 
to publication. Valuable comments from an anonymous referee also significantly 
improved the  quality and clarity of the paper.

%%%%%%%%%%%%%%%%%%%%%%%%%%%%%%%%%%%%%%%%%%%%%%%%%%



\bibliographystyle{mnras}
\bibliography{gaia} % if your bibtex file is called example.bib

%%%%%%%%%%%%%%%%%%%%%%%%%%%%%%%%%%%%%%%%%%%%%%%%%%

%%%%%%%%%%%%%%%%% APPENDICES %%%%%%%%%%%%%%%%%%%%%

%%%%%%%%%%%%%%%%%%%%%%%%%%%%%%%%%%%%%%%%%%%%%%%%%%

% Don't change these lines
\bsp	% typesetting comment
\label{lastpage}
\end{document}